\documentclass[manuscript]{acmart}

\AtBeginDocument{%
  \providecommand\BibTeX{{%
    \normalfont B\kern-0.5em{\scshape i\kern-0.25em b}\kern-0.8em\TeX}}}

\acmDOI{none}

\usepackage{nccmath}
\usepackage{multirow}
\usepackage{subcaption}
\usepackage{arydshln}
\usepackage{color}

\begin{document}

\title[RECipe: A Multi-Modal Recipe KG Fits Multi-Purpose Recommendation]{RECipe: Does a Multi-Modal Recipe Knowledge Graph Fit a Multi-Purpose Recommendation System?}

\author{Ali Pesaranghader}
\authornote{Corresponding author: ali.pesaranghader@lge.com}
\email{ali.pesaranghader@lge.com}

\author{Touqir Sajed}
\email{touqir.sajed@lge.com}


\affiliation{
  \institution{LG Electronics, Toronto AI Lab}
  \city{Toronto}
  \state{Ontario}
  \country{Canada}
}

\renewcommand{\shortauthors}{Pesaranghader and Sajed}

\begin{abstract}
Over the past two decades, recommendation systems (RSs) have used machine learning (ML) solutions to recommend items, e.g., movies, books, and restaurants, to clients of a business or an online platform. Recipe recommendation, however, has not yet received much attention compared to those applications. 
We introduce RECipe as a multi-purpose \textit{recipe} recommendation framework with a multi-modal knowledge graph (MMKG) backbone. The motivation behind RECipe is to go beyond (deep) neural collaborative filtering (NCF) by recommending recipes to users when they query in natural language or by providing an image. RECipe consists of 3 subsystems: (1) behavior-based recommender, (2) review-based recommender, and (3) image-based recommender. Each subsystem relies on the embedding representations of entities and relations in the graph. 
We first obtain (pre-trained) embedding representations of textual entities, such as reviews or ingredients, from a fine-tuned model of Microsoft's MPNet. We initialize the weights of the entities with these embeddings to train our knowledge graph embedding (KGE) model.
For the visual component, i.e., recipe images, we develop a KGE-Guided variational autoencoder (KG-VAE) to learn the distribution of images and their latent representations. 
Once KGE and KG-VAE models are fully trained, we use them as a \textit{multi-purpose} recommendation framework. For benchmarking, we created two knowledge graphs (KGs) from public datasets on Kaggle for recipe recommendation. Our experiments show that the KGE models have comparable performance to the neural solutions. We also present pre-trained NLP embeddings to address important applications such as \textit{zero-shot inference} for new users (or the cold start problem) and \textit{conditional recommendation} with respect to recipe categories. We eventually demonstrate the application of RECipe in a multi-purpose recommendation setting. 
\end{abstract}



\acmConference{}{}{}
\setcopyright{none}
\settopmatter{printacmref=false} 
\renewcommand\footnotetextcopyrightpermission[1]{} 
\fancyfoot{}
\maketitle


\section{Introduction}

Online and e-commerce platforms, e.g., Netflix, Amazon and Tripadvisor, have benefited from recommendation systems (RSs) to recommend items, e.g., movies, books, clothes, or restaurants, to their clients for the past two decades, if not longer. 
Recipe recommendation, however, has not yet received much attention compared to those domains by the research community, despite the fact that there are uncountable resources for various foods, recipes, and cuisines on social media such as Instagram\footnote{https://www.instagram.com/explore/tags/recipes/}, Pinterest\footnote{https://www.pinterest.ca/natashaskitchen/the-most-popular-recipes-on-pinterest/}, and other platforms like Food.com\footnote{https://www.food.com/} and Allrecipes.com\footnote{https://www.allrecipes.com/}. 
Such abundance is implicitly indicative of the interest in as well as the need for recipe recommendation systems. Predictably, food recommendation systems will eventually become an inseparable part of our daily life. In other words, RSs will help us search for recipes, walk us through their instructions, and prepare a meal while they consider our personal preferences, diet and health. They can even potentially assist professional chefs in serving their clients better than ever. Therefore, the importance of developing recipe recommendation solutions is undebatable.

We focus on three RS settings in this work\footnote{We overlook other settings, such as session-based recommendation, because they plausibly fall into the main three groups described above. Our focus is instead on the input modality types.}: (1) behavior-based recommendation\footnote{Also known as link-based recommendation in the knowledge graph community.}, (2) text-based recommendation, and (3) image-based recommendation. 
The behavior-based algorithms learn (implicit or explicit) patterns from users' past behaviours, e.g., \textit{viewed}, \textit{purchased}, or \textit{liked}, to recommend new items. Content-based and collaborative filtering, as examples of behavior-based solutions, are commonly used in the RS applications \cite{duricic2018trust, fan2019deep, yang2018mmcf, zhu2018automatic, bogaards2021content, wang2019overview, lops2019trends, 10.5555/2457524.2457626}.
A text-based recommender retrieves relevant items by matching an input query, or a profile, with the textual information of items, e.g., news titles or movie descriptions \cite{9354169, 6927564, 10.1145/1864708.1864756, 8349947, zheng2017joint}. Review-based solutions, as a subdiscipline of text-based RS, benefit from valuable information in (textual) reviews for personalization and recommendation \cite{10.1007/s11257-015-9155-5, 10.1145/3018661.3018665, 10.1145/3430984.3431069, 10.1145/3397271.3401281, 10.1145/3383313.3412207, 10.1145/3308560.3316601, 10.1145/3097983.3098170}.
For image-based recommendation, a user may upload an image with the intention of item recognition and retrieval of similar items, potentially along with their (descriptive) information \cite{10.1145/2964284.2973834, 10.1145/3038912.3052638, 10.1145/3159652.3159728, 10.1145/2766462.2767755, bossard2014food, marin2019recipe1m+}. Recall that our primary interest is \textbf{\textit{recipe}} recommendation; therefore, we explain the foundation of our work for that domain hereafter.

To the best of our knowledge, there has been no solution that handles the aforementioned recommendation settings holistically in one place, particularly for recommending recipes. 
Knowledge graphs (KGs) are a potential solution to connect all these tasks with different modalities (e.g., structured data, texts, and images) for multi-purpose recommendation. We consider an RS as a multi-purpose solution if it accomplishes behavior-based, text-based, and image-based recommendations together\footnote{We do not use the multi-task terminology because the main task is still recommendation regardless of the (input) modalities.}. 
In this work, we introduce RECipe as a multi-purpose \textit{recipe} recommendation system that benefits from KGs with multi-modalities\footnote{It is also known as \textit{cross-modalities} in the literature.}. We summarize our key contributions below.

\begin{enumerate}
    \item We introduce RECipe as  a multi-purpose \textit{recipe} recommendation solution.
    \item We present two knowledge graphs (KGs) for recipe recommendation or retrieval. The KGs can be considered as benchmarks for future research work.
    \item We address zero-shot inference for new users (or the cold start problem) by employing and aligning pre-trained NLP embeddings to obtain their initial embedding representations.
    \item We introduce conditional recommendation with respect to categories of recipes to improve overall accuracy considering various ranking measures.
    \item We present the RECipe applications in behavior-based, review-based, and image-based recommendations.
\end{enumerate}

The remainder of this paper is organized as follows. We study related works to the recipe recommendation or retrieval tasks in Section \ref{sec:related-works}. We present our RECipe framework and its components in Section \ref{sec:recipe-framework}. In Section \ref{sec:experimental-evaluation}, we define our research questions, then conduct extensive experiments for the multi-purpose recommendations. Finally, we conclude the paper and discuss potential future work in Section \ref{sec:conclusion-and-future-work}.


\section{Related Works}
\label{sec:related-works}

In this section, we study related works for \textit{recipe} recommendation (or retrieval) for \textit{behavior-based}, \textit{text-based}, and \textit{image-based} settings.

\begin{itemize}
    \item \textit{behavior-based solutions} refer to methods that use the contents of items or users' preferences, either implicit or explicit, for recommendation.
    \citet{forbes2011content} proposed a content-boosted matrix factorization algorithm for incorporating the content information as a natural linear constraint into the matrix factorization (MF) algorithm for collaborative filtering. They showed that their algorithm could provide informative insights about the contents that would, otherwise, not be available for the recipe recommendation task.
    \citet{teng2012recipe} claimed that they predict recipe ratings with features derived from the ingredient networks and nutrition facts. Further, they presented that their ingredient network made ingredient substitution feasible.
    \item \textit{Text-based solutions} may benefit from textual information such as recipe names, descriptions, instructions, or user profiles or reviews for recipe recommendation or retrieval tasks. Although the NLP methods were used in settings such as ingredient networks \cite{teng2012recipe}, Recipe1M+ \cite{marin2019recipe1m+}, and pFoodREQ \cite{chen2021personalized}, we found no \textit{fully} text-based query-recipe and profile-recipe matching solutions in the literature.
    \item \textit{Image-based solutions} use image processing models to categorize recipe pictures and retrieve information for a given image. \citet{bossard2014food} introduced the Food-101 dataset for visual recipe classification, where every recipe belongs to one of 101 categories. They used Random Forests to mine discriminative components and applied a superpixel-based patch sampling strategy for classifying recipe images.
    \citet{chen2016deep} presented another dataset containing 65K Chinese recipes with 110K images. The focus of their work was not only food categorization but also ingredient recognition. They developed deep architectures to simultaneously learn ingredient recognition and food categorization by exploiting the mutual but also fuzzy relationship among them. They later extended their work in \cite{chen2017cross} to extract ingredients, cutting and cooking methods from an image query for retrieving recipes.
    M\textsuperscript{3}TDBN \cite{min2016being} used a multi-modal multi-task beep belief network to learn joint image-ingredient representation for recipe classification/retrieval and ingredient inference from food images.
    Recipe1M+ \cite{marin2019recipe1m+} is a large-scale and structured dataset with over 1M cooking recipes and 13M food images. The authors trained a neural network to learn a joint embedding of pictures and recipes for the image-recipe retrieval task. Their experiment showed that the solution could map images to recipes and vice versa.
\end{itemize}

We also briefly describe knowledge base (KB) and KG-based related works to food recommendation below.

\begin{itemize}
    \item \textit{KG-based solutions} use graphical algorithms or representations for the recommendation tasks. The focus of most works is news, book or movie recommendations \cite{10.1145/3292500.3330989, 10.1145/3240323.3240361, 10.1145/3308558.3313411, polignano2021together}. For example, a recent work by \citet{polignano2021together} used knowledge graph embeddings and (deep) contextual embeddings for book and movie recommendations. 
    On the other hand, there are a handful of contributions to recipe recommendation. \citet{haussmann2019foodkg} introduced FoodKG, a large-scale and unified food knowledge graph (KG), which brings together food ontologies, recipes, ingredients and nutritional data.
    By relying on FoodKG, pFoodREQ \cite{chen2021personalized} formulates food recommendation as a constrained question answering task over a large-scale food knowledge base (KBQA). In addition to the requirements from the user query, personalized requirements from the user’s dietary preferences and health guidelines are handled in a unified way as additional constraints to the QA system.
\end{itemize}

We summarize the drawbacks of existing methods as follows. No pure text-based, as well as review-based, solution exists for recipe recommendation. Recall that, a review-based recommender retrieves relevant items by matching an input query, e.g., \textit{"I am looking for spicy Thai noodles"} or \textit{"A perfect meal for a date"}, with other people’s reviews written for the recipes. Reviewers’ comments potentially make a rich corpus that holds indirect and crowdsourced information about recipes such as tips, ingredient replacements, or occasions that are not available with recipes' (textual) information. Most image-based solutions were used for recipe categorization than the recommendation tasks. Besides, only a few considered ingredient recognition as well. None of the existing datasets or knowledge graphs have users’ comments/reviews. Finally, there is a lack of \textit{stand-alone} solutions for fulfilling behavior-based, review-based, and image-based \textit{recipe} recommendations together.

To address the challenges above, we introduce the RECipe framework as a holistic, i.e., 3-in-1, solution for behavior-based, review-based, and image-based recommendation. 
We describe RECipe and the KGs in Sections \ref{sec:recipe-framework} and \ref{sec:experimental-evaluation}, respectively.

\begin{figure*}[t]
    \centering
    \resizebox{0.95\linewidth}{!}{
        \includegraphics{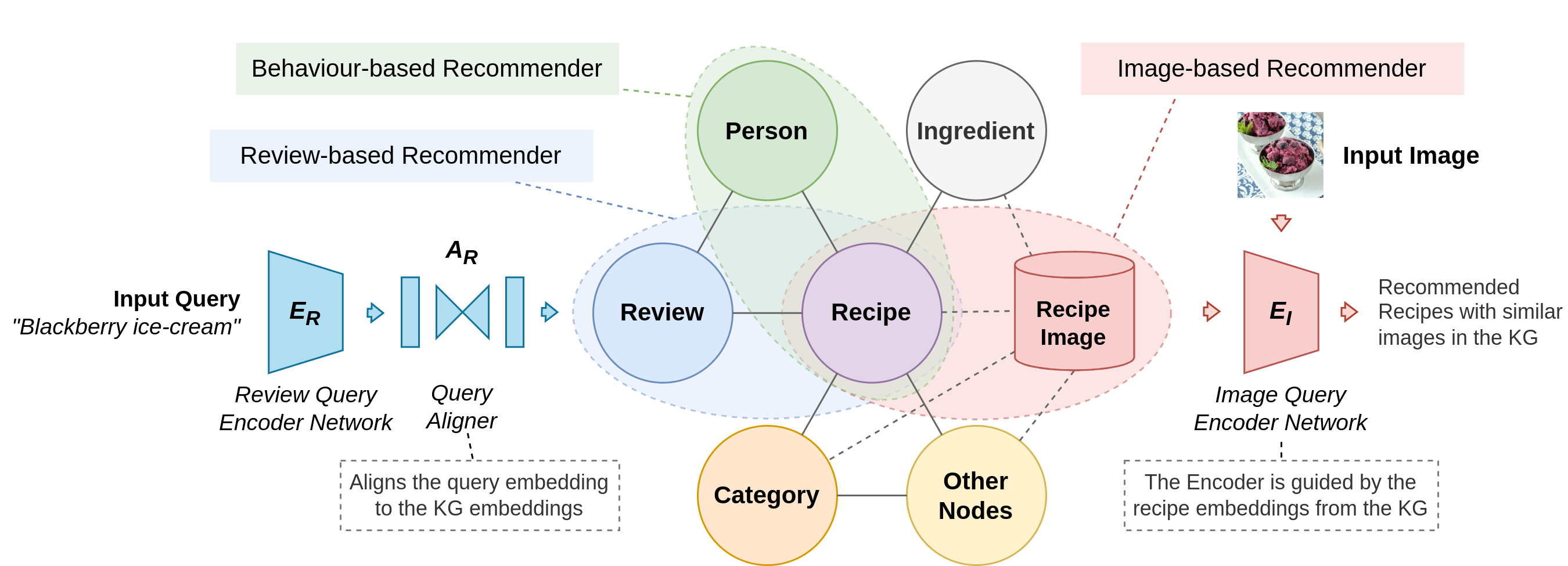}
    }
    \caption{The RECipe Framework}
    \label{fig:RECipe}
\end{figure*}

\section{RECipe Framework}
\label{sec:recipe-framework}

We present the RECipe framework and its components in Figure \ref{fig:RECipe}. The heart of RECipe is a knowledge graph with entities such as recipes, persons\footnote{We interchangeably use \textit{person} and \textit{user} in this paper.}, ingredients amongst others. RECipe uses knowledge graph embedding (KGE) models to learn entity and relation embeddings for the recommendation task. The framework holds three subsystems of \textit{behavior-based}, \textit{review-based}, and \textit{image-based recommenders}. The behavior-based recommender is defined as a link prediction task between the person and recipe entities in the graph. That is, the recommender predicts whether a user likes a recipe. The review-based recommender encodes an input query, e.g., \textit{``Blackberry ice-cream''}, and matches it with reviews of all recipes for recommendation.
Likewise, the image-based recommender obtains the latent representation of an input image for finding similar recipes or foods. We discuss how the entity weights are initialized before describing every component in the following subsections.

\begin{figure*}[h]
    \centering
    \resizebox{0.7\linewidth}{!}{\includegraphics{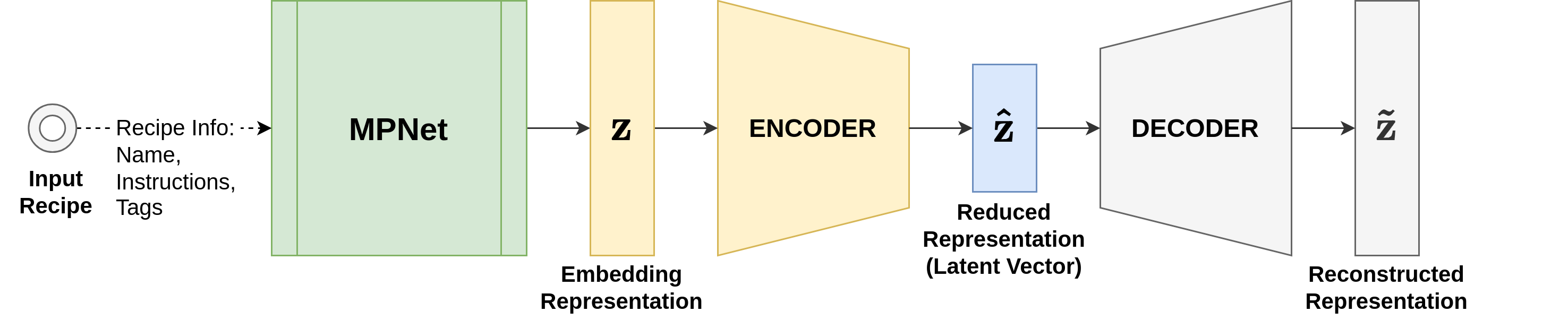}}
    \caption{MPNet + AutoEncoder (Learning Reduced Representation of Textual Entities for Entity Weight Initialization)}
    \label{fig:mp-net-ae}
\end{figure*}

\noindent \textit{Entity Weight Initialization.} The entity weights may be initialized by either random embeddings or (pre-trained) NLP embeddings before training/fine-tuning our KGE models. 
The NLP embeddings could be obtained from the Statistical-based Embedding methods \cite{levy2014neural, pesaranghader2019deepbiowsd}, the word representation models such as Word2Vec \cite{mikolov2013efficient} and GloVe \cite{pennington-etal-2014-glove}, and language models (LMs) such as BERT \cite{devlin-etal-2019-bert}, RoBERTa \cite{DBLP:journals/corr/abs-1907-11692} and MPNet \cite{10.5555/3495724.3497138, reimers-gurevych-2019-sentence}. We, in particular, employ the neural NLP embeddings obtained from a fine-tuned MPNet model\footnote{Sentence Transformers: \href{https://huggingface.co/sentence-transformers/all-mpnet-base-v2}{all-mpnet-base-v2} (We chose this model because of it promising performance across different tasks.)} whenever entity initialization with pre-trained embedding is considered. We may also reduce the dimension of embedding representations using an AutoEncoder to relax memory bottlenecks if required. Figure \ref{fig:mp-net-ae} illustrates how an AutoEncoder reduces the dimension of the embeddings obtained by the MPNet model.

We obtain a neural NLP embedding for each entity using their (textual) properties. For example, Eq.\ \eqref{eq:neural-emb} shows how an embedding vector is calculated for a given recipe, person, or review entity. Function $z$ embeds a textual input\footnote{Recall that we use MPNet for this function.}.
For recipe $k$, denoted by $rcp_k$, the embedding vector is the average of its name and instructions embedding vectors. The embedding vector for person $k$, indicated by $psn_k$, is the average of embedding vectors of their reviews, i.e., $rvw_1$ to $rvw_m$ (where $m$ is the total number of reviews written by person $k$), in the training data. We may replace $\textbf{z}$ with $\hat{\textbf{z}}$ when we reduce the dimension of the embeddings. 
Similarly, we obtain embeddings of ingredients and (recipe) categories. Once all pre-trained embeddings are obtained, we initialize the entity weights with them for KGE training for the corresponding settings/experiments.

\begin{equation}
    \label{eq:neural-emb}
    \begin{gathered}
        \textbf{z}_{rcp_k} = ((z(rcp_k^{name}) + z(rcp_k^{instructions}))/{2}  \\
        \textbf{z}_{psn_k} = \frac{1}{m} \sum^{m}_{i=1} \left(\textbf{z}_{rvw_i}\right) 
        \mbox{{ }{}{where}{ }{ }} \textbf{z}_{rvw_i} = z(rvw_i^{content}) \mbox{{ }{for}{ }{$psn_k$}} \\         
    \end{gathered}
\end{equation}

\subsection{behavior-based Recommender}

The behavior-based recommendation system (BRS) relies on the links among the person and recipe entities in the graph.
Recall that, if a user likes a recipe, there is a link between them.
The \textit{Person-Recipe} subgraph is shown by a green ellipse in Figure \ref{fig:RECipe}. 
We train a KGE model, e.g., RotatE \cite{sun2019rotate}, to predict whether a link exists between a user and a recipe. 
The score function $f$ measures the salience of a candidate triplet $(h, r, t)$ and its formulation varies from one KGE algorithm to another.
In our setting, regardless of the KGE model, the function $f$ receives the embeddings of users and recipes as well as the embedding of the predicate (or relation) to calculate their triplet scores. Eq.\ \eqref{equ:score_function} calculates the score for triplet $(psn_i, p_{psn:likes:rcp}, rcp_j)$:
\begin{equation}
\label{equ:score_function}
    score = f(\textbf{z}_{psn_i}, \textbf{z}_{psn:likes:rcp}, \textbf{z}_{rcp_j})
\end{equation}

\noindent where $\textbf{z}_{psn_i}$, $\textbf{z}_{psn:likes:rcp}$, and $\textbf{z}_{rcp_j}$ are the embeddings of user $i$, predicate \textit{psn:likes:rcp}, and recipe $j$, respectively. The optimization goal is to give a higher score to \textit{true} triplet $(h, r, t)$ than corrupted ones, i.e., $(h', r, t)$ or $(h, r, t')$.
We then use the trained KGE model for link prediction, and then rank predictions w.r.t. their scores for recommendation.

\subsection{Review-based Recommender}
\label{subsec:rvw-recom}

\begin{figure*}[h]
    \centering
    \includegraphics[width=12cm]{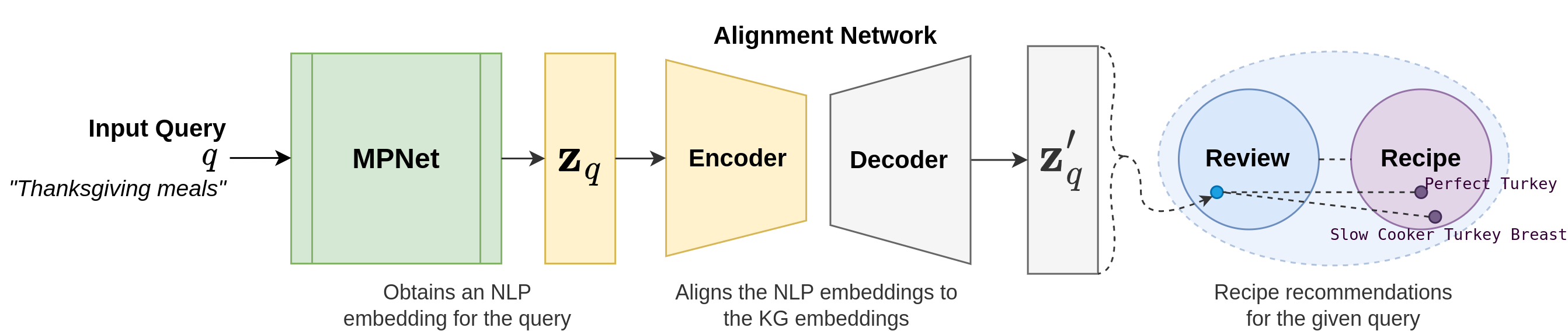}
    \caption{Review-based Recommender with Alignment Network}
    \label{fig:review-based-recommender}
\end{figure*}

The goal of a review-based recommendation system (RRS) is to recommend recipes related to a (natural language) query, e.g., \textit{``A great birthday cake for my son!''} or \textit{``Thanksgiving meals''}. Figure \ref{fig:review-based-recommender} shows how our review-based recommender operates. 
Initially, we train our KGE model to predict whether there is a relation between a review and a recipe. Next, for query $q$, we also train an \textit{alignment} network to align NLP embeddings ($\textbf{z}_qs$) to KGE embeddings of reviews ($\textbf{z}_q^\prime s$), as shown in Figure \ref{fig:review-based-recommender}. This step is necessary for inference because we need to align a query to the KG embedding latent space. During the inference step, we obtain an aligned representation for an input query to recommend recipes via link prediction. For example, \textit{``Perfect Turkey''} and \textit{``Slow Cooker Turkey Breast''} are recommended for query \textit{``Thanksgiving meals''}.

\subsection{Image-based Recommender}
\label{subsec:kg-guided-vae}

\begin{figure*}[!]
    \centering
    \resizebox{0.85\linewidth}{!}{\includegraphics{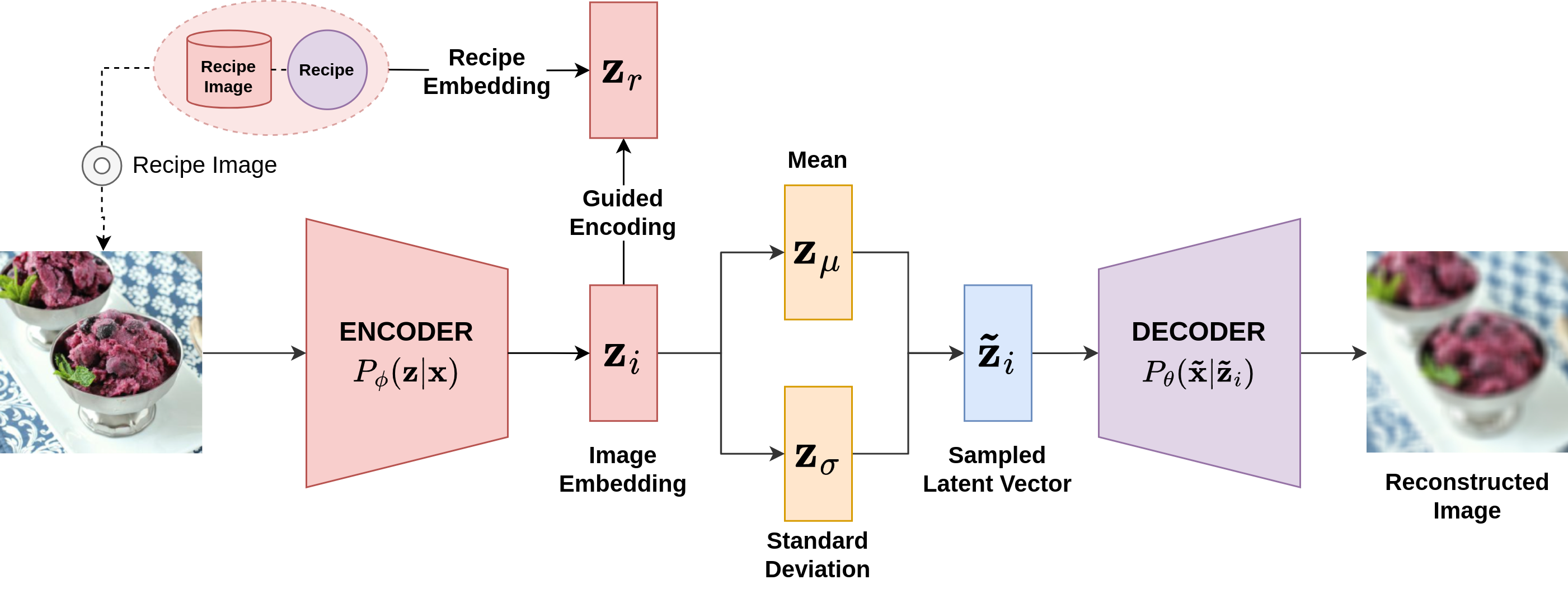}}
    \caption{Architecture of KGE-Guided VAE (KG-VAE) to Learn Representations of Visual Entities (i.e., Recipe Images)}
    \label{fig:guided-vae}
\end{figure*}

The image-based recommendation system (IRS) identifies and retrieves similar recipes to a given recipe image. The core of our IRS is a KGE-guided variational autoencoder (KG-VAE) in which the VAE \cite{kingma2013auto} is guided to the KG embedding space while it learns the distribution of images. Figure \ref{fig:guided-vae} shows the architecture of KG-VAE. During training, the VAE is guided with the KG embeddings of recipes to force images from a similar recipe family to be closer to one another. For that, we use the loss function in Eq.\ \eqref{equ:kge-guided-vae-loss-function} where $\lambda$ is a hyperparameter to control the contribution of VAE's ELBO, $\textbf{z}_i$ and $\textbf{z}_b$ are image and recipe embeddings, and MSE is the mean squared error loss function.

\begin{equation}
    \label{equ:kge-guided-vae-loss-function}
    \begin{gathered}
        \mathcal{L}_\textit{KG-VAE} = \mathcal{L}_\textit{KGE-Guidance} - \lambda \times ELBO_\textit{VAE} \\
        \mathcal{L}_\textit{KGE-Guidance} = MSE(\textbf{z}_i, \textbf{z}_r)
    \end{gathered}
\end{equation}

During the inference, the encoder part is only used to encode a given image. Then, the encoded representation is compared with the representations of other recipe images to retrieve similar recipes.

\section{Experimental Evaluation}
\label{sec:experimental-evaluation}

In this section, we first describe our research questions and then present two recipe KGs along with our experimental settings. Finally, we present and discuss our experimental results for each research question.

\subsection{Research Questions}

Our experiments investigate five research questions that are defined as follows. The reader should note that these research questions reflect our main contributions.

\begin{itemize}
    \item \textbf{RQ1 - KGE-based Recommendation vs. Neural Recommendation.} 
    Do KGE algorithms perform comparably to neural solutions for the recipe recommendation task?
    \item \textbf{RQ2 - Pre-trained Neural NLP Embeddings for Zero-shot Inference.}
    Does weight initialization of KG entities with pre-trained neural NLP embeddings improve link prediction for recipe recommendation for new users?
    \item \textbf{RQ3 - Complementary Sub-graphs.} Would the inclusion of additional explicit (e.g., ingredients) and implicit (e.g., latent categories) entities incorporated in the KG augment the system?
    \item \textbf{RQ4 - Conditional Recommendation.} Why is conditional recipe recommendation necessary, and how does the system benefit from decoupling a user's preferences?
    \item \textbf{RQ5 - Multi-modal KG Fits Multi-Purpose RS.} Could a multi-modal KG effectively realize a multi-purpose system, e.g., review-based recommendation, and image-based recommendation?
\end{itemize}

\subsection{Experimental Design}
\label{subsec:experimental-design}

\noindent \textbf{Datasets.} For the purpose of this research work, we need datasets that contain information related to recipes, including, names, instructions, and ingredients as well as users' ratings and reviews. In that regard, we consider the food.com and allrecipes.com datasets, publicly available on Kaggle.com, for our research work\footnote{Food.com: \url{https://www.kaggle.com/irkaal/foodcom-recipes-and-reviews}}\textsuperscript{,}\footnote{Allrecipes.com: \url{https://www.kaggle.com/elisaxxygao/foodrecsysv1}}\textsuperscript{,}\footnote{We use the datasets only for non-commercial research purpose.}:

\begin{itemize}
    \item \textbf{Food.com} is a digital brand featuring recipes from home cooks and celebrity chefs, food news, and pop culture. The food.com Kaggle dataset holds records about recipes, e.g., name, instructions, categories, and ingredients as well as users' ratings and reviews. We pre-processed the data and extracted entities and relationships to build the Food.com KG. 
    Table \ref{tab:foodcom} summarizes entities and relations in the KG.
    \item \textbf{Allrecipes.com} is an online social networking platform for publishing, rating, and reviewing recipes.
    The Kaggle dataset provides us with recipes, reviews, and their interactions. We pre-processed the data and transformed it into the Allrecipes.com KG. Table \ref{tab:allrecipes} lists entities and relations in the Allrecipes.com KG.
\end{itemize}

In this study, we consider reviews with a rating of 4 or 5 as \textit{positive} only for the purpose of recommendation.

\noindent \textbf{Artifacts.} 
We used PyKeen\footnote{https://pykeen.github.io/} for the KGE models, and NeuRec\footnote{https://github.com/wubinzzu/NeuRec} from NExT++\footnote{https://github.com/NExTplusplus/NeuRec} for the (neural) baselines for RQ1.

\noindent \textbf{KGE Model.} We use RotatE \cite{sun2019rotate} as the KGE model to train embeddings of the entities and relations in our knowledge graphs. RotatE models relations as rotations from head to tail entities in complex space as follows $\textbf{z}_t = \textbf{z}_h \odot \textbf{r}_r$ where $\textbf{z}, \textbf{r} \in \mathbb{C} ^{d}$ and the complex elements of $\textbf{r}_r$ are restricted to have a modulus of one, i.e., $||\textbf{r}_r|| = 1$. We train RotatE with the original Self-adversarial Negative Sampling (NSSA) loss function.
We also evaluated other KGE models, such as TransE \cite{bordes2013translating}, TransH \cite{Wang_Zhang_Feng_Chen_2014}, and DistMult \cite{DBLP:journals/corr/YangYHGD14a}, in our preliminary experiments and none of them showed comparable results to RotatE. As a result, we consider RotatE as our base KGE model in RECipe.

\noindent \textbf{(KGE) Hyperparameters.} We fine-tune hyperparameters on the \emph{validation sets} in a grid search way. The ranges of hyperparameters are as follows: embedding dimension $d \in \{32, 64, 128, 256, 512, 768\}$, learning rate $\alpha \in \{0.01, 0.001, 0.0001\}$, number of negative samples $n^- \in \{1, 5, 10, 50, 100\}$, and margin $\gamma \in \{1, 5, 10, 20\}$. We also use Adam \cite{DBLP:journals/corr/KingmaB14} to optimize the NSSA loss function.

\noindent \textbf{Evaluation Metrics.} The ranking measures are used to reward recommendation lists where relevant items are placed on the top. The measures are \textit{hit rate (HR)}, \textit{normalized discounted cumulative gain (nDCG)}, \textit{mean reciprocal ranking (MRR)}, and \textit{mean rank (MR)}. Those measures are widely used in the RS and KGE literature \cite{he2017neural, polignano2021together}. We use the \emph{binary} relevance version of nDCG since our recommendation task is binary (likes/exists or not) \cite{wang2013theoretical}.

\subsection{RQ1: KGE-based Recommendation vs. Neural Recommendation}
\label{subsec:rq1}

We compare RotatE and the neural solutions only against the \textit{Person-User} sub-graphs (as a bipartite graph) in our two recipe KGs.

\noindent \textit{Neural Models (Baselines).} MLP \cite{he2017neural} is a multi-layer perceptron that receives the feature vectors of items and users as inputs to predict whether there is an interaction between them. NeuMF \cite{he2017neural} is an item-based collaborative filtering solution that learns the interactions among users and items by fusing MLP with generalized matrix factorization (GMF) algorithms. CDAE \cite{wu2016collaborative} uses denoising auto-encoders (DAE) to learn the embeddings of users and items for top-K recommendation. CFGAN \cite{chae2018cfgan} is a GAN-based collaborative filtering framework that utilizes vector-wise adversarial training for taking full advantage of adversarial training to improve the recommendation accuracy. JCA \cite{10.1145/3308558.3313678} jointly trains two auto-encoders to learn both user-user and item-item correlations to recover the user-item rating matrix.


\vspace{1mm}

\noindent \textit{(Baselines) Hyperparameters.} 
We run each algorithm 200 epochs for hyperparameter tuning against the validation set. We report the best results on the test sets.

\vspace{1mm}

\noindent \textit{Data Split.} By following a series of works such as \cite{he2017neural}, we used the \textit{leave-one-out} (LOO) approach to split triplets with relation type \textit{psn:likes:rcp} into training and test sets. LOO ensures that we have one random (\textit{psn:likes:rcp}) interaction per person in our test data.

\vspace{1mm}

\noindent \textit{Experimental Results.} Table \ref{tab:kge-vs-neural} presents the results of our experiments for RQ1. 
It is seen that RotatE outperforms the neural models in terms of the ranking measures. The performance of RotatE degrades when the embedding dimension $d$ decreases from 768 to 64; however, RotatE$_{d=64}$ yet outperforms the neural algorithms while showing comparable results to CDAE. 
CDAE has better performance than the other neural models with a notable margin which could be explained by the presence of the user node in its architecture.
Finally, the Hit@10, nDCG@10, MRR@10 results confirm that the task of recipe recommendation is much more challenging than the studied domains. Therefore, we consider this work as one of the first few contributions to recipe recommendation which may attract the research community.

\begin{table*}[htp]
    \centering
    \begin{tabular}{c|c|c|c|c|c|c|}
        \cline{2-7}
         & \multicolumn{3}{c|}{Food.com} & \multicolumn{3}{c|}{Allrecipes.com} \\ \hline
        \multicolumn{1}{|l|}{Algorithm} & Hit@10 & nDCG@10 & MRR@10 & Hit@10 & nDCG@10 & MRR@10\\ \hline
        \multicolumn{1}{|l|}{MLP} & 0.04640 & 0.02385 & 0.01707 & 0.03459 & 0.01777 & 0.01272 \\
        \multicolumn{1}{|l|}{NeuMF} & 0.05067 & 0.02540 & 0.01779 & 0.03677 & 0.01811 & 0.01253 \\
        \multicolumn{1}{|l|}{CDAE} & 0.05805 & 0.03117 & 0.02309 & 0.04057 & 0.02039 & 0.01431 \\
        \multicolumn{1}{|l|}{CFGAN} & 0.04349 & 0.02319 & 0.01711 & 0.03631 & 0.01745 & 0.01176 \\
        \multicolumn{1}{|l|}{JCA} & 0.04834 & 0.02391 & 0.01664 & 0.03333 & 0.01560 & 0.01034 \\ \hline \hline
        \multicolumn{1}{|l|}{RotatE$_{d=64}$} & 0.05863 & 0.02885 & 0.02109 & 0.0385 & 0.01822 & 0.01276 \\
        \multicolumn{1}{|l|}{RotatE$_{d=256}$} & 0.06174 & 0.03149 & 0.02353 & 0.04103 & 0.01991 & 0.01450 \\
        \multicolumn{1}{|l|}{RotatE$_{d=768}$} & \textbf{0.06504} & \textbf{0.03326} & \textbf{0.02525} & \textbf{0.04298} & \textbf{0.02082} & \textbf{0.01499} \\ \hline
    \end{tabular}
    \caption{Comparison of KGE-based and Neural-based RS Algorithms}
    \label{tab:kge-vs-neural}
\vspace{-6mm}    
\end{table*}

Next, we investigate how the initialization of entity weights with the pre-trained NLP embeddings improves the prediction power for the zero-shot recommendation. 

\subsection{RQ2: Pre-trained Neural NLP Embeddings for Zero-shot Inference}

\begin{figure*}[ht]
    \centering
    \resizebox{0.95\linewidth}{!}{\includegraphics{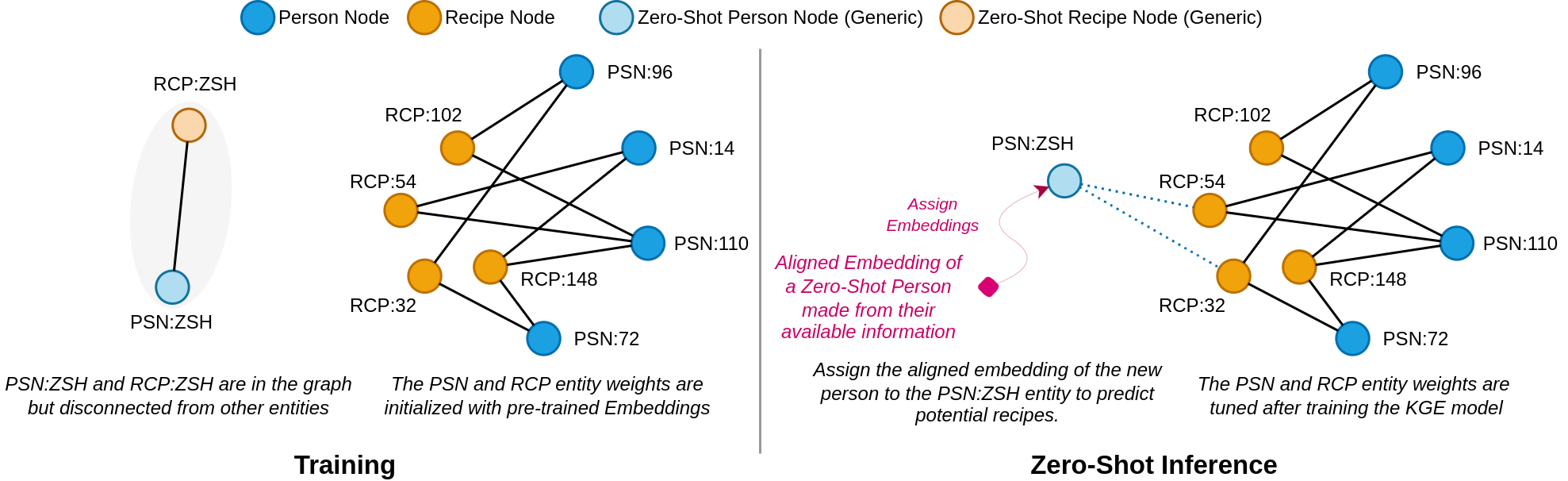}}
    \caption{Illustration of Zero-Shot Inference for Recipe Recommendation to New Users}
    \label{fig:zeroshot-structure}
\end{figure*}

The \emph{cold start} problem refers to a well-known RS challenge when new users join a platform, and no recommendation can be made for them because they have no rating nor few interactions, e.g., clicks or views. We refer to this problem as \textit{zero-shot inference} for new users, and we show how to settle this problem by employing (pre-trained) NLP embeddings in this section. Fig.\ \ref{fig:zeroshot-structure} is the illustration of our solution for zero-shot inference. First, we add a placeholder node \textit{PSN:ZSH} to the KG and keep it disconnected from all (real) recipes. We, then, train our KGE model to learn interactions among (existing) users and recipes. For zero-shot inference/recommendation for a new user, we obtain an NLP embedding representation from their ``available/immediate'' information from their profile, preferences, diet habits, etc. 
Then, we align the NLP embedding to a KG embedding (similar to what we have in Fig.\ \ref{fig:review-based-recommender}). Finally, we assign the aligned embedding to the \textit{PSN:ZSH} node and predict recipes via link prediction.

\vspace{1mm}

\noindent \textit{Data Splits.} We first extract some users to create a \textit{zero-shot} holdout set. Those users will act as \textit{new} users for our evaluation and they are disconnected from all recipes during training. To make the task challenging, we consider users that have relations with ``\textit{isolated}'' recipes. Otherwise, recommending the popular recipes is not difficult and the task may suffer from \textit{popularity bias}. Since the reviews are a good representative of users' preferences, we use them to obtain the NLP embeddings for the new users.
Then, the remaining \textit{person-recipe} interactions are split into training, validation, and testing sets. We ensure that the size of validation and test sets are equal to that of the zero-shot holdout set. We split the data $10$ times with random seeds and report the average results of our experiments.

\vspace{1mm}

\noindent \textit{Experimental Results.} Table \ref{tab:nlp-zeroshot} shows the results of our experiments for zero-shot inference. In the table, we compare three options for embedding assignment for a new user: (1) a random embedding (RAND) from the embedding space of other users, (2) the average of the embeddings of existing users (AVG), and (3) the NLP and then KG-aligned embedding (KG-aligned). We see that the RotatE\textsubscript{\textit{KG-aligned}} outperforms two other options. The difference with RotatE\textsubscript{\textit{RAND}} is much more significant. It is noteworthy that RotatE\textsubscript{\textit{KG-aligned}} not only outperforms RotatE\textsubscript{\textit{AVG}} but also leads to better-personalized recommendations because RotatE\textsubscript{\textit{AVG}} relies on the average of embeddings of other users.

\begin{table*}[htp]
    \centering
    \resizebox{\linewidth}{!}{
    \begin{tabular}{c|c|c|c|c|c|c|c|c|}
        \cline{2-9}
         & \multicolumn{4}{c|}{Food.com} & \multicolumn{4}{c|}{Allrecipes.com} \\ \cline{2-9}
         & \multicolumn{2}{c|}{$K=10$} & \multicolumn{2}{c|}{$\infty$} & \multicolumn{2}{c|}{$K=10$} & \multicolumn{2}{c|}{$\infty$} \\ \hline
        \multicolumn{1}{|l|}{Algorithm} & Hit & nDCG & MRR & MR & Hit & nDCG & MRR & MR \\ \hline
        \multicolumn{1}{|l|}{} & 0.01060 \textsuperscript{$\star$} & 0.00488 \textsuperscript{$\star$} & 0.00738 \textsuperscript{$\star$} & 935.1 \textsuperscript{$\star$} & 0.00724 \textsuperscript{$\star$} & 0.00325 \textsuperscript{$\star$} & 0.00541 \textsuperscript{$\star$} & 1676.1 \textsuperscript{$\star$} \\ [-3pt]
        \multicolumn{1}{|l|}{\multirow{-2}{*}{RotatE$_{d=64\textit{+RAND}}$}} & \footnotesize $\pm$ 0.00085 & \footnotesize $\pm$ 0.00048 & \footnotesize $\pm$ 0.00042 & \footnotesize $\pm$ 8.82 & \footnotesize $\pm$ 0.00050 & \footnotesize $\pm$ 0.00026 & \footnotesize $\pm$ 0.00024 & \footnotesize $\pm$ 13.22 \\ \hdashline
        \multicolumn{1}{|l|}{} & 0.01021 \textsuperscript{$\star$} & 0.00472 \textsuperscript{$\star$} & 0.00761 \textsuperscript{$\star$} & 909.1 \textsuperscript{$\star$} & 0.00724 \textsuperscript{$\star$} & 0.00323 \textsuperscript{$\star$} & 0.00563 \textsuperscript{$\star$} & 1628.6 \textsuperscript{$\star$} \\ [-3pt]
        \multicolumn{1}{|l|}{\multirow{-2}{*}{RotatE$_{d=256\textit{+RAND}}$}} & \footnotesize $\pm$ 0.00079 & \footnotesize $\pm$ 0.00040 & \footnotesize $\pm$ 0.00035 & \footnotesize $\pm$ 6.26 & \footnotesize $\pm$ 0.00055 & \footnotesize $\pm$ 0.00033 & \footnotesize $\pm$ 0.00033 & \footnotesize $\pm$ 19.24 \\ \hdashline
        \multicolumn{1}{|l|}{} & 0.01044 \textsuperscript{$\star$} & 0.00475 \textsuperscript{$\star$} & 0.00764 \textsuperscript{$\star$} & 904.1 \textsuperscript{$\star$} & 0.00722 \textsuperscript{$\star$} & 0.00322 \textsuperscript{$\star$} & 0.00565 \textsuperscript{$\star$} & 1621.3 \textsuperscript{$\star$} \\ [-3pt]
        \multicolumn{1}{|l|}{\multirow{-2}{*}{RotatE$_{d=768\textit{+RAND}}$}} & \footnotesize $\pm$ 0.00082 & \footnotesize $\pm$ 0.00038 & \footnotesize $\pm$ 0.00030 & \footnotesize $\pm$ 10.05 & \footnotesize $\pm$ 0.00055 & \footnotesize $\pm$ 0.00029 & \footnotesize $\pm$ 0.00033 & \footnotesize $\pm$ 15.20 \\ \hline
        \multicolumn{1}{|l|}{} & 0.02360 \textsuperscript{$\star$} & 0.01164 \textsuperscript{$\star$} & 0.01316 \textsuperscript{$\diamond$} & 859.7 \textsuperscript{$\star$} & 0.01830 \textsuperscript{$\star$} & 0.00872 \textsuperscript{$\star$} & 0.01001 \textsuperscript{$\star$} & 1565.1 \textsuperscript{$\star$} \\ [-3pt]
        \multicolumn{1}{|l|}{\multirow{-2}{*}{RotatE$_{d=64\textit{+AVG}}$}} & \footnotesize $\pm$ 0.00177 & \footnotesize $\pm$ 0.00088 & \footnotesize $\pm$ 0.00074 & \footnotesize $\pm$ 7.50 & \footnotesize $\pm$ 0.00113 & \footnotesize $\pm$ 0.00058 & \footnotesize $\pm$ 0.00050 & \footnotesize $\pm$ 16.98 \\ \hdashline
        \multicolumn{1}{|l|}{} & 0.02335 \textsuperscript{$\star$} & 0.01154 \textsuperscript{$\star$} & 0.01310 \textsuperscript{$\diamond$} & 859.9 \textsuperscript{$\star$} & 0.01824 \textsuperscript{$\star$} & 0.00870 \textsuperscript{$\star$} & 0.01002 \textsuperscript{$\star$} & 1565.3 \textsuperscript{$\star$} \\ [-3pt]
        \multicolumn{1}{|l|}{\multirow{-2}{*}{RotatE$_{d=256\textit{+AVG}}$}} & \footnotesize $\pm$ 0.0015 & \footnotesize $\pm$ 0.00078 & \footnotesize $\pm$ 0.00069 & \footnotesize $\pm$ 7.64 & \footnotesize $\pm$ 0.00109 & \footnotesize $\pm$ 0.00050 & \footnotesize $\pm$ 0.00047 & \footnotesize $\pm$ 17.50 \\ \hdashline
        \multicolumn{1}{|l|}{} & 0.02362 \textsuperscript{$\star$} & 0.01164 \textsuperscript{$\diamond$} & 0.01318 \textsuperscript{$\diamond$} & 862.4 \textsuperscript{$\star$} & 0.01834 \textsuperscript{$\star$} & 0.00869 \textsuperscript{$\star$} & 0.00994 \textsuperscript{$\star$} & 1570.7 \textsuperscript{$\star$} \\ [-3pt]
        \multicolumn{1}{|l|}{\multirow{-2}{*}{RotatE$_{d=768\textit{+AVG}}$}} & \footnotesize $\pm$ 0.00160 & \footnotesize $\pm$ 0.00083 & \footnotesize $\pm$ 0.00075 & \footnotesize $\pm$ 7.62 & \footnotesize $\pm$ 0.00118 & \footnotesize $\pm$ 0.00057 & \footnotesize $\pm$ 0.00047 & \footnotesize $\pm$ 16.96 \\ \hline \hline
        \multicolumn{1}{|l|}{} & 0.02521 \textsuperscript{$\star$} & 0.01175 \textsuperscript{$\star$} & 0.01323 & 818.7 \textsuperscript{$\star$} & 0.01981 \textsuperscript{$\diamond$} & 0.00948 \textsuperscript{$\star$} & 0.01083 \textsuperscript{$\star$} & 1476.3 \textsuperscript{$\diamond$} \\ [-3pt]
        \multicolumn{1}{|l|}{\multirow{-2}{*}{RotatE$_{d=64\textit{+KG-aligned}}$}} & \footnotesize $\pm$ 0.00224 & \footnotesize $\pm$ 0.00119 & \footnotesize $\pm$ 0.00105 & \footnotesize $\pm$ 13.33 & \footnotesize $\pm$ 0.00167 & \footnotesize $\pm$ 0.00078 & \footnotesize $\pm$ 0.00056 & \footnotesize $\pm$ 26.02 \\ \hdashline
        \multicolumn{1}{|l|}{} & \textbf{0.02888} & \textbf{0.01387} & \textbf{0.01519} & \textbf{780.3} & \textbf{0.02393} & \textbf{0.01163} & \textbf{0.01275} & \textbf{1420.5} \\ [-3pt]
        \multicolumn{1}{|l|}{\multirow{-2}{*}{RotatE$_{d=256\textit{+KG-aligned}}$}} & \footnotesize $\pm$ 0.00208 & \footnotesize $\pm$ 0.00120 & \footnotesize $\pm$ 0.00120 & \footnotesize $\pm$ 15.21 & \footnotesize $\pm$ 0.00228 & \footnotesize $\pm$ 0.00110 & \footnotesize $\pm$ 0.00102 & \footnotesize $\pm$ 30.70 \\ 
        \hdashline
        \multicolumn{1}{|l|}{} & 0.02757 & 0.01330 & 0.01472 & 803.3 \textsuperscript{$\diamond$} & 0.02310 & 0.01128 & 0.01255 & 1428.0 \\ [-3pt]
        \multicolumn{1}{|l|}{\multirow{-2}{*}{RotatE$_{d=768\textit{+KG-aligned}}$}} & \footnotesize $\pm$ 0.00174 & \footnotesize $\pm$ 0.00090 & \footnotesize $\pm$ 0.00075 & \footnotesize $\pm$ 8.00 & \footnotesize $\pm$ 0.00174 & \footnotesize $\pm$ 0.00089 & \footnotesize $\pm$ 0.00077 & \footnotesize $\pm$ 15.88 \\ \hline
    \end{tabular}
    }
    \caption{Pre-trained NLP and KG-Aligned Embeddings for Zero-shot Learning ($\infty$ is for all items ranked.) The symbols $\diamond$ and $\star$ show that there is a significant difference between the corresponding setting and the best setting (shown in bold), statistically validated by the Wilcoxon signed rank test with $\mbox{p-value} < 0.005$ and $\mbox{p-value} < 0.001$, respectively.}
    \label{tab:nlp-zeroshot}
\vspace{-4mm}   
\end{table*}

\subsection{RQ3: Complementary Sub-graphs}

We now study and evaluate the impact of adding other sub-graphs to the KGs for recipe recommendation. We consider \textit{Recipe-Category}, \textit{Recipe-Ingredient}, \textit{Ingredient-Ingredient} sub-graphs for Food.com and \textit{Recipe-Ingredient} and \textit{Ingredient-Ingredient} sub-graphs for Allrecipes.com. Further, we derive \textit{latent recipe clusters} and include them in our graphs. We explain how we find the latent recipe clusters below.

\vspace{1mm}

\noindent \textit{Latent Recipe Clusters.} Allrecipes.com and Food.com data neither come with nor have rich and informative (recipe) category entity types. We, therefore, add another entity type to our graphs based on the recipe clusters. For that purpose, we used K-means to cluster recipes where each recipe is represented by its neural NLP embedding. We ran K-Means with different random seeds and found the best number of clusters based on the Elbow method against the sum of squared distances (SSD) of samples to their closest cluster center. We also used Silhouette Coefficient \cite{ROUSSEEUW198753} to confirm the results of the Elbow method. We finally derived $44$ and $52$ well-separated clusters for the Food.com and Allrecipes.com recipes, respectively. 



Once the clusters are extracted, we add the \textit{recipe:belongs-to:recipe-cluster} and \textit{person:relates-to:recipe-cluster} sub-graphs to our KGs. For pre-trained entity weight initialization, the embedding of each cluster is the mean of the (pre-trained) embeddings of its recipes. It is worth mentioning that this experiment shows another application of (pre-trained) NLP embeddings that helps us enhance our knowledge graphs by adding the \textit{recipe clusters} entity type. The clusters also lead to \textit{conditional recipe recommendation} that we cover in Section \ref{subsec:rq4-conditional-recommendation}.

\vspace{1mm}

\noindent \textit{Data Split.} We keep 80\% of the \textit{user-recipe} triplets for training, 10\% for validation, and another 10\% for testing. We split the data with different random seeds 10 times and then average the results. The triplets of sub-graphs are added to the training set each time.

\vspace{1mm}

\noindent \textit{Experimental Results.} Table \ref{tab:kge-subgraphs} shows that adding sub-graphs (SG) improved the overall performance results. Hit@10 improved by 6{\textperthousand} and $1\%$ for Food.com and Allrecipes.com, respectively. The mean rank (MR) improvement is also noticeable. 
In overall, the improvement is more significant for Allrecipes.com which may be explained by the nature of its data and how entities interact with another.


\begin{table*}[h!]
    \centering
    \begin{tabular}{c|c|c|c|c|c|c|c|c|}
        \cline{2-9}
         & \multicolumn{4}{c|}{Food.com} & \multicolumn{4}{c|}{Allrecipes.com} \\ \cline{2-9}
         & \multicolumn{2}{c|}{$K=10$} & \multicolumn{2}{c|}{$\infty$} & \multicolumn{2}{c|}{$K=10$} & \multicolumn{2}{c|}{$\infty$} \\ \hline
        \multicolumn{1}{|l|}{Algorithm} & Hit & nDCG & MRR & MR & Hit & nDCG & MRR & MR \\ \hline
        \multicolumn{1}{|l|}{} & 0.04087 \textsuperscript{$\star$} & 0.01998 \textsuperscript{$\star$} & 0.02080 \textsuperscript{$\star$} & 719.7 \textsuperscript{$\star$} & 0.02764 \textsuperscript{$\star$} & 0.01305 \textsuperscript{$\star$} & 0.01425 \textsuperscript{$\star$} & 1239.6 \textsuperscript{$\star$} \\ [-3pt]
        \multicolumn{1}{|l|}{\multirow{-2}{*}{RotatE$_{d=64}$}} & \footnotesize $\pm$ 0.00284 & \footnotesize $\pm$ 0.00181 & \footnotesize $\pm$ 0.00173 & \footnotesize $\pm$ 9.18 & \footnotesize $\pm$ 0.00231 & \footnotesize $\pm$ 0.00114 & \footnotesize $\pm$ 0.00107 & \footnotesize $\pm$ 35.10 \\ \hdashline
        \multicolumn{1}{|l|}{} & 0.05063 \textsuperscript{$\star$} & 0.02502 \textsuperscript{$\star$} & 0.02564 \textsuperscript{$\star$} & 621.6 \textsuperscript{$\star$} & 0.03321 \textsuperscript{$\star$} & 0.01615 \textsuperscript{$\star$} & 0.01720 \textsuperscript{$\star$} & 1186.7 \textsuperscript{$\star$} \\ [-3pt]
        \multicolumn{1}{|l|}{\multirow{-2}{*}{RotatE$_{d=256}$}} & \footnotesize $\pm$ 0.00304 & \footnotesize $\pm$ 0.00161 & \footnotesize $\pm$ 0.00137 & \footnotesize $\pm$ 9.31 & \footnotesize $\pm$ 0.00516 & \footnotesize $\pm$ 0.00241 & \footnotesize $\pm$ 0.00209 & \footnotesize $\pm$ 47.74 \\ \hdashline
        \multicolumn{1}{|l|}{} & 0.05289 \textsuperscript{$\star$} & 0.02631 \textsuperscript{$\star$} & 0.02677 \textsuperscript{$\star$} & 611.4 \textsuperscript{$\star$} & 0.03741 \textsuperscript{$\star$} & 0.01811 \textsuperscript{$\star$} & 0.01901 \textsuperscript{$\star$} & 1133.5 \textsuperscript{$\star$} \\ [-3pt]
        \multicolumn{1}{|l|}{\multirow{-2}{*}{RotatE$_{d=768}$}} & \footnotesize $\pm$ 0.00334 & \footnotesize $\pm$ 0.00193 & \footnotesize $\pm$ 0.00176 & \footnotesize $\pm$ 11.16 & \footnotesize $\pm$ 0.00228 & \footnotesize $\pm$ 0.00114 & \footnotesize $\pm$ 0.00092 & \footnotesize $\pm$ 33.51 \\ \hline \hline
        \multicolumn{1}{|l|}{} & 0.04354 \textsuperscript{$\diamond$} & 0.02105 \textsuperscript{$\star$} & 0.02176 \textsuperscript{$\star$} & 664.9 \textsuperscript{$\star$} & 0.03435 \textsuperscript{$\star$} & 0.01634 \textsuperscript{$\star$} & 0.01761 \textsuperscript{$\star$} & 1107.8 \textsuperscript{$\star$} \\ [-3pt]
        \multicolumn{1}{|l|}{\multirow{-2}{*}{RotatE$_{d=64\textit{+SG}}$}} & \footnotesize $\pm$ 0.00547 & \footnotesize $\pm$ 0.00267 & \footnotesize $\pm$ 0.00233 & \footnotesize $\pm$ 43.75 & \footnotesize $\pm$ 0.00591 & \footnotesize $\pm$ 0.00279 & \footnotesize $\pm$ 0.00246 & \footnotesize $\pm$ 103.86 \\ \hdashline
        \multicolumn{1}{|l|}{} & 0.05829 & 0.02859 & 0.02879 & \textbf{548.8} & 0.04569 & 0.02143 & 0.02192 & \textbf{967.2} \\ [-3pt]
        \multicolumn{1}{|l|}{\multirow{-2}{*}{RotatE$_{d=256\textit{+SG}}$}} & \footnotesize $\pm$ 0.00369 & \footnotesize $\pm$ 0.00180 & \footnotesize $\pm$ 0.00148 & \footnotesize $\pm$ 7.98 & \footnotesize $\pm$ 0.00401 & \footnotesize $\pm$ 0.00197 & \footnotesize $\pm$ 0.00154 & \footnotesize $\pm$ 22.95 \\ \hdashline
        \multicolumn{1}{|l|}{} & \textbf{0.05873} & \textbf{0.02899} & \textbf{0.02932} & 551.5 \textsuperscript{$\diamond$} & \textbf{0.04619} & \textbf{0.02191} & \textbf{0.02235} & 981.0 \textsuperscript{$\star$} \\ [-3pt]
        \multicolumn{1}{|l|}{\multirow{-2}{*}{RotatE$_{d=768\textit{+SG}}$}} & \footnotesize $\pm$ 0.00365 & \footnotesize $\pm$ 0.00208 & \footnotesize $\pm$ 0.00198 & \footnotesize $\pm$ 8.24 & \footnotesize $\pm$ 0.00398 & \footnotesize $\pm$ 0.00190 & \footnotesize $\pm$ 0.00180 & \footnotesize $\pm$ 25.27 \\ \hline
    \end{tabular}
    \caption{Complementary Sub-graphs (SG) ($\infty$ is for all items ranked.)The symbols $\diamond$ and $\star$ show that there is a significant difference between the corresponding setting and the best setting (shown in bold), statistically validated by the Wilcoxon signed rank test with $\mbox{p-value} < 0.005$ and $\mbox{p-value} < 0.001$, respectively.}
    \label{tab:kge-subgraphs}
\vspace{-4mm}
\end{table*}

Next, we study conditional recommendation with respect to the clusters that we derived in this experiment.

\subsection{RQ4: Conditional Recommendation}
\label{subsec:rq4-conditional-recommendation}

\begin{figure*}[htp]
    \centering
    \resizebox{0.85\linewidth}{!}{\includegraphics{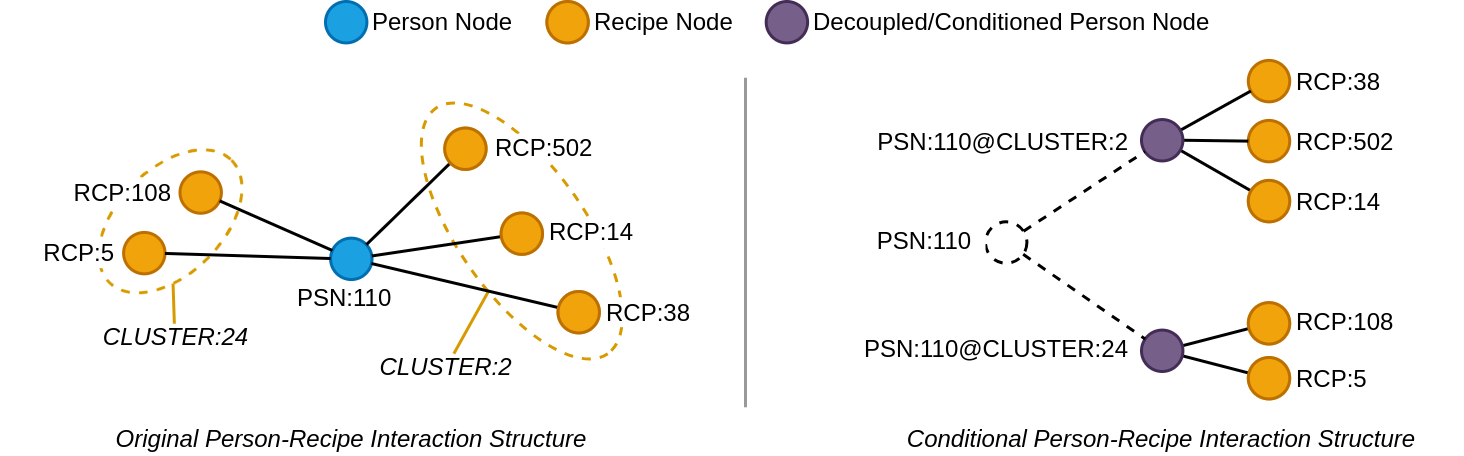}}
    \caption{Original vs. Conditional Person-Recipe Interaction Structures}
    \label{fig:cr-and-decoupling}
\end{figure*}

We introduce conditional recommendation (CR) based on the clusters of recipes in this section. The CR setting conditions the task of recommendation on the recipe clusters that a person may like. That is, we decouple the person entity nodes to the \textit{person@recipe-cluster} entity nodes in the graph. Figure \ref{fig:cr-and-decoupling} visualizes how we decouple a person entity node with respect to its corresponding recipe clusters. In this example, person \textit{PSN:110} likes recipes \textit{RCP:14}, \textit{RCP:38}, and \textit{RCP:502} from recipe cluster \textit{CLUSTER:2} as well as recipes \textit{RCP:5} and \textit{RCP:108} from recipe cluster \textit{CLUSTER:24} in the \textit{original person-recipe interaction} structure. On the other hand, in the \textit{conditional person-recipe interaction} structure, the person node is decoupled to two conditional person nodes \textit{PSN:110@CLUSTER:2} and \textit{PSN:110@CLUSTER:24} while they are still connected to the recipes they like as in the original structure. The intuition behind conditional recommendation is to avoid confusing a recommender when users like very disjoint recipes (items). The conditional recommendation is another task that the pre-trained NLP embeddings made feasible.

\noindent \textit{Data Splits.} We split our data as follows: $80\%$ of the \textit{person-recipe} interactions goes for training, $10\%$ for validation, and $10\%$ for testing. The validation and testing sets hold the same triplets for the Non-CR and CR settings with the difference that the person nodes are decoupled in CR. The sub-graphs' triplets are also added to the training set each time.

\vspace{1mm}

\noindent \textit{Experimental Results.} Table \ref{tab:noncond-vs-cond} compares the results of Non-CR and CR. It is seen that the CR setting significantly improved the recommendation results for all evaluation measures. For example, Hit@10 increased by $37\%$ and $30\%$ for Food.com and Allrecipes.com, respectively. The mean rank (MR) also substantially improved after conditioning our recommendations.


\begin{table*}[htp]
    \centering
    \begin{tabular}{c|c|c|c|c|c|c|c|c|}
        \cline{2-9}
         & \multicolumn{4}{c|}{Food.com} & \multicolumn{4}{c|}{Allrecipes.com} \\ \cline{2-9}
         & \multicolumn{2}{c|}{$K=10$} & \multicolumn{2}{c|}{$\infty$} & \multicolumn{2}{c|}{$K=10$} & \multicolumn{2}{c|}{$\infty$} \\ \hline
        \multicolumn{1}{|l|}{Algorithm} & Hit & nDCG & MRR & MR & Hit & nDCG & MRR & MR \\ \hline
        \multicolumn{1}{|l|}{} & 0.04573 \textsuperscript{$\star$} & 0.02215 \textsuperscript{$\star$} & 0.02260 \textsuperscript{$\star$} & 684.9 \textsuperscript{$\star$} & 0.03813 \textsuperscript{$\star$} & 0.01834 \textsuperscript{$\star$} & 0.01905 \textsuperscript{$\star$} & 1113.3 \textsuperscript{$\star$} \\ [-3pt]
        \multicolumn{1}{|l|}{\multirow{-2}{*}{RotatE$_{d=64}$}} & \footnotesize $\pm$ 0.00306 & \footnotesize $\pm$ 0.00147 & \footnotesize $\pm$ 0.00113 & \footnotesize $\pm$ 26.44 & \footnotesize $\pm$ 0.00524 & \footnotesize $\pm$ 0.00246 & \footnotesize $\pm$ 0.00224 & \footnotesize $\pm$ 97.23 \\ \hdashline
        \multicolumn{1}{|l|}{} & 0.05942 \textsuperscript{$\star$} & 0.02936 \textsuperscript{$\star$} & 0.02967 \textsuperscript{$\star$} & 552.4 \textsuperscript{$\star$} & 0.04721 \textsuperscript{$\star$} & 0.02312 \textsuperscript{$\star$} & 0.02334 \textsuperscript{$\star$} & 999.9 \textsuperscript{$\star$} \\ [-3pt]
        \multicolumn{1}{|l|}{\multirow{-2}{*}{RotatE$_{d=256}$}} & \footnotesize $\pm$ 0.00128 & \footnotesize $\pm$ 0.00086 & \footnotesize $\pm$ 0.00077 & \footnotesize $\pm$ 6.79 & \footnotesize $\pm$ 0.00153 & \footnotesize $\pm$ 0.00083 & \footnotesize $\pm$ 0.00075 & \footnotesize $\pm$ 13.14 \\ \hdashline
        \multicolumn{1}{|l|}{} & 0.06168 \textsuperscript{$\star$} & 0.03063 \textsuperscript{$\star$} & 0.03078 \textsuperscript{$\star$} & 547.7 \textsuperscript{$\star$} & 0.05062 \textsuperscript{$\star$} & 0.0247 \textsuperscript{$\star$} & 0.02468 \textsuperscript{$\star$} & 991.1 \textsuperscript{$\star$} \\ [-3pt]
        \multicolumn{1}{|l|}{\multirow{-2}{*}{RotatE$_{d=768}$}} & \footnotesize $\pm$ 0.00213 & \footnotesize $\pm$ 0.00116 & \footnotesize $\pm$ 0.00105 & \footnotesize $\pm$ 8.51 & \footnotesize $\pm$ 0.00178 & \footnotesize $\pm$ 0.00111 & \footnotesize $\pm$ 0.00096 & \footnotesize $\pm$ 10.90 \\ \hline \hline
        \multicolumn{1}{|l|}{} & 0.32505 \textsuperscript{$\star$} & 0.15780 \textsuperscript{$\star$} & 0.13867 \textsuperscript{$\star$} & 43.9 \textsuperscript{$\star$} & 0.24260 \textsuperscript{$\star$} & 0.11598 \textsuperscript{$\star$} & 0.10300 \textsuperscript{$\star$} & 174.5 \textsuperscript{$\star$} \\ [-3pt]
        \multicolumn{1}{|l|}{\multirow{-2}{*}{RotatE$_{d=64\textit{+CR}}$}} & \footnotesize $\pm$ 0.00512 & \footnotesize $\pm$ 0.00303 & \footnotesize $\pm$ 0.00246 & \footnotesize $\pm$ 9.39 & \footnotesize $\pm$ 0.00551 & \footnotesize $\pm$ 0.00348 & \footnotesize $\pm$ 0.00258 & \footnotesize $\pm$ 89.42 \\ \hdashline
        \multicolumn{1}{|l|}{} & 0.33369 \textsuperscript{$\star$} & 0.16355 \textsuperscript{$\star$} & 0.14420 \textsuperscript{$\star$} & 36.1 \textsuperscript{$\star$} & 0.25326 \textsuperscript{$\star$} & 0.12200 \textsuperscript{$\star$} & 0.10944 \textsuperscript{$\star$} & 100.6 \\ [-3pt]
        \multicolumn{1}{|l|}{\multirow{-2}{*}{RotatE$_{d=256\textit{+CR}}$}} & \footnotesize $\pm$ 0.00523 & \footnotesize $\pm$ 0.00427 & \footnotesize $\pm$ 0.00388 & \footnotesize $\pm$ 8.52 & \footnotesize $\pm$ 0.00565 & \footnotesize $\pm$ 0.00364 & \footnotesize $\pm$ 0.00307 & \footnotesize $\pm$ 52.10 \\ \hdashline
        \multicolumn{1}{|l|}{} & \textbf{0.43216} & \textbf{0.22838} & \textbf{0.20445} & \textbf{24.0} & \textbf{0.35835} & \textbf{0.18941} & \textbf{0.17064} & \textbf{55.8} \\ [-3pt]
        \multicolumn{1}{|l|}{\multirow{-2}{*}{RotatE$_{d=768\textit{+CR}}$}} & \footnotesize $\pm$ 0.00291 & \footnotesize $\pm$ 0.00258 & \footnotesize $\pm$ 0.00321 & \footnotesize $\pm$ 1.05 & \footnotesize $\pm$ 0.00317 & \footnotesize $\pm$ 0.0023 & \footnotesize $\pm$ 0.00251 & \footnotesize $\pm$ 7.60 \\ \hline
    \end{tabular}
    \caption{Non-conditional vs.\ Conditional Recommendation (CR) ($\infty$ is for all items ranked.) The symbols $\diamond$ and $\star$ show that there is a significant difference between the corresponding setting and the best setting (shown in bold), statistically validated by the Wilcoxon signed rank test with $\mbox{p-value} < 0.005$ and $\mbox{p-value} < 0.001$, respectively.}
    \label{tab:noncond-vs-cond}
\vspace{-6mm}    
\end{table*}


\subsection{RQ5: Multi-modal KG Fits Multi-Purpose RS}

\subsubsection{Review-based Recommendation} We present the results of our review-based recommendation systems (RRS) in this section. We compare three solutions, namely, MPNet RRS, KGE RRS, and Hybrid RRS, as listed in Table \ref{tab:review-based-systems}.

\begin{itemize}
    \item MPNet RRS uses the MPNet model \cite{song2020mpnet, reimers-gurevych-2019-sentence} to embed an input query as well as reviews. It then compares their embeddings using the cosine similarity measure. The recipes with the most similar reviews to the query are then recommended.
    \item KGE RRS is a trained RotatE\textsubscript{$d=768$} model against each KG with the Review-Recipe subgraph included. As explained in Section \ref{subsec:rvw-recom}, we embed an input query using the MPNet model and then align it to the embedding space of the KGE model using an aligner network. Recall that, an aligner network projects the MPNet embeddings to the KG embedding space. Then, the aligned query embedding is used as a review (KG) embedding to predict potential/relative recipes.
    \item Hybrid RRS combines the recipe ranking of MPNet RRS and KGE RRS for recipe recommendation. That is, the rank of a recipe is the average of its ranks with MPNet RRS and KGE RSS.
\end{itemize}

\noindent \textit{Query Creation.} In the absence of real queries paired with relevant recipes, we hold out a set of reviews from both KGs and modify them to represent such queries.

\vspace{1mm}

\noindent \textit{Experimental Results.} Table \ref{tab:review-based-systems} presents the results of our experiments for review-based recommendation. MPNet RRS led to better results compared to KGE RRS. This can be explained by the fact that the MPNet model is trained and fine-tuned for various NLP-related tasks. Therefore, it shows a better performance than KGE RRS. It may, however, miss some latent facts among reviews and recipes which KGE RRS may potentially learn and capture. Hybrid RSS benefits from the strength of the two solutions to better rank recipes. Hybrid RRS shows a significant performance compared to MPNet RSS and KGE RSS with respect to all evaluation measures for both data. Finally, we present a few real examples of the (Hybrid) RRS results in Table \ref{tab:rrs-examples}. You will be redirected to a recipe's webpage if you click on its icon (e.g., \includegraphics[width=0.3cm,height=0.3cm,keepaspectratio]{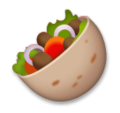}).


\begin{table*}[htp]
    \centering
    \begin{tabular}{c|c|c|c|c|c|c|c|c|}
        \cline{2-9}
         & \multicolumn{4}{c|}{Food.com} & \multicolumn{4}{c|}{Allrecipes.com} \\ \cline{2-9}
         & \multicolumn{2}{c|}{$K=10$} & \multicolumn{2}{c|}{$\infty$} & \multicolumn{2}{c|}{$K=10$} & \multicolumn{2}{c|}{$\infty$} \\ \hline
        \multicolumn{1}{|l|}{Algorithm} & Hit & nDCG & MRR & MR & Hit & nDCG & MRR & MR \\ \hline

        \multicolumn{1}{|l|}{} & 0.36415 \textsuperscript{$\diamond$} & 0.21642 & \textbf{0.20382} & 198.6 \textsuperscript{$\star$} & 0.25225 \textsuperscript{$\star$} & 0.14162 \textsuperscript{$\star$} & 0.13051 \textsuperscript{$\star$} & 437.7 \textsuperscript{$\star$} \\ [-3pt]
        \multicolumn{1}{|l|}{\multirow{-2}{*}{MPNet RRS}} & \footnotesize $\pm$ 0.03680 & \footnotesize $\pm$ 0.02535 & \footnotesize $\pm$ 0.02333 & \footnotesize $\pm$ 29.47 & \footnotesize $\pm$ 0.01258 & \footnotesize $\pm$ 0.01011 & \footnotesize $\pm$ 0.00877 & \footnotesize $\pm$ 31.83 \\ \hdashline
        \multicolumn{1}{|l|}{} & 0.30726 \textsuperscript{$\star$} & 0.16388 \textsuperscript{$\star$} & 0.14117 \textsuperscript{$\star$} & 202.7 \textsuperscript{$\star$} & 0.23687 \textsuperscript{$\star$} & 0.13604 \textsuperscript{$\star$} & 0.12854 \textsuperscript{$\star$} & 425.2 \textsuperscript{$\star$} \\ [-3pt]
        \multicolumn{1}{|l|}{\multirow{-2}{*}{KGE RRS}} & \footnotesize $\pm$ 0.03902 & \footnotesize $\pm$ 0.02643 & \footnotesize $\pm$ 0.02232 & \footnotesize $\pm$ 24.96 & \footnotesize $\pm$ 0.01396 & \footnotesize $\pm$ 0.01124 & \footnotesize $\pm$ 0.01074 & \footnotesize $\pm$ 30.89 \\ \hline \hline

        \multicolumn{1}{|l|}{} & \textbf{0.39950} & \textbf{0.21823} & 0.19059 & \textbf{159.8} & \textbf{0.29876} & \textbf{0.17487} & \textbf{0.16394} & \textbf{342.1} \\ [-3pt]
        \multicolumn{1}{|l|}{\multirow{-2}{*}{Hybrid RRS}} & \footnotesize $\pm$ 0.03886 & \footnotesize $\pm$ 0.02242 & \footnotesize $\pm$ 0.02232 & \footnotesize $\pm$ 19.50 & \footnotesize $\pm$ 0.01327 & \footnotesize $\pm$ 0.01173 & \footnotesize $\pm$ 0.01029 & \footnotesize $\pm$ 18.96 \\ \hline
    \end{tabular}
    \caption{Review-based Recommendation Solutions (RRS) ($\infty$ is for all items ranked.) The symbols $\diamond$ and $\star$ show that there is a significant difference between the corresponding setting and the best setting (shown in bold), statistically validated by the Wilcoxon signed rank test with $\mbox{p-value} < 0.05$ and $\mbox{p-value} < 0.01$, respectively.}
    \label{tab:review-based-systems}
\vspace{-4mm}    
\end{table*}


\begin{table*}[htp]
    \captionsetup{justification=centering}
    
    \begin{tabular}{|l|p{5cm}|p{6.75cm}|}
    \hline
    KG  & Query & Top 3 Recipes \\ \hline
    \multirow{6}{*}{Food.com}       
        & ``\textit{Recommend meatball sandwiches with whole grain buns and mozzarella cheese}'' & 
            Kittencal's Italian Melt-in-your-mouth Meatballs \href{https://www.food.com/recipe/kittencals-italian-melt-in-your-mouth-meatballs-69173}{\includegraphics[width=0.3cm,height=0.3cm,keepaspectratio]{img/stuffed-flatbread.png}} \newline 
            Divine Meatball Sandwiches \href{https://www.food.com/recipe/divine-meatball-sandwiches-59289}{\includegraphics[width=0.3cm,height=0.3cm,keepaspectratio]{img/stuffed-flatbread.png}} \newline 
            Mama Iuliucci's Famous Meat-a-balls \href{https://www.food.com/recipe/mama-iuliuccis-famous-meat-a-balls-italian-meatballs-32618}{\includegraphics[width=0.3cm,height=0.3cm,keepaspectratio]{img/stuffed-flatbread.png}} \\ \cdashline{2-3} 
        & ``\textit{Enchiladas with cottage cheese or sour cream}'' &
            Easiest Beef Enchiladas Ever! \href{https://www.food.com/recipe/easiest-beef-enchiladas-ever-17586}{\includegraphics[width=0.3cm,height=0.3cm,keepaspectratio]{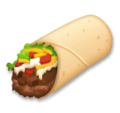}} \newline
            Cottage Cheese Enchiladas \href{https://www.food.com/recipe/cottage-cheese-enchiladas-49595}{\includegraphics[width=0.3cm,height=0.3cm,keepaspectratio]{img/burrito.png}} \newline
            Chicken Enchiladas \href{https://www.food.com/recipe/chicken-enchiladas-24041}{\includegraphics[width=0.3cm,height=0.3cm,keepaspectratio]{img/burrito.png}}
    \\ \hline
    \multirow{6}{*}{Allrecipes.com} 
        & ``\textit{The best lemonade or watermelon margaritas}'' & 
            Beer Margaritas \href{https://www.allrecipes.com/recipe/63191/beer-margaritas/}{\includegraphics[width=0.3cm,height=0.3cm,keepaspectratio]{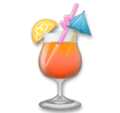}} \newline 
            Sangria! Sangria! \href{https://www.allrecipes.com/recipe/72612/sangria-sangria/}{\includegraphics[width=0.3cm,height=0.3cm,keepaspectratio]{img/drink.png}} \newline 
            Jewel's Watermelon Margaritas \href{https://www.allrecipes.com/recipe/221229/jewels-watermelon-margaritas/}{\includegraphics[width=0.3cm,height=0.3cm,keepaspectratio]{img/drink.png}} \\ \cdashline{2-3} 
        & ``\textit{Chicken salads with mayo}'' & 
            Basic Chicken Salad \href{https://www.allrecipes.com/recipe/8499/basic-chicken-salad/}{\includegraphics[width=0.3cm,height=0.3cm,keepaspectratio]{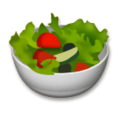}} \newline
            Strawberry Spinach Salad I \href{https://www.allrecipes.com/recipe/14276/strawberry-spinach-salad-i/}{\includegraphics[width=0.3cm,height=0.3cm,keepaspectratio]{img/salad.png}} \newline
            Fruity Curry Chicken Salad \href{https://www.allrecipes.com/recipe/8932/fruity-curry-chicken-salad/}{\includegraphics[width=0.3cm,height=0.3cm,keepaspectratio]{img/salad.png}}
        \\ \hline
    \end{tabular}
    \caption{Real Application of Hybrid Review-based Recommendation (H-RRS)}
    \label{tab:rrs-examples}
\vspace{-4mm}
\end{table*}

\subsubsection{Image-Based Recommendation}

We use our KGE-Guided VAE (KG-VAE) for the image-based recommendation. Recall that, Section \ref{subsec:kg-guided-vae} introduces KG-VAE and explains how we train and use it for inference.

\vspace{1mm}

\noindent \textit{Data Split.} We split our data into training, validation, and test sets w.r.t. the \textit{Recipe-Ingredient} relation in this task. That relation potentially enforces the KGE model to learn the latent relations among recipes and ingredients. Once the recipe embeddings are learned, we use them to train our KG-VAE model. We fine-tune hyperparameters, e.g., learning rate and $\lambda$, on a validation set before the inference task.



\vspace{1mm}

\begin{figure*}[t]
    \resizebox{\textwidth}{!}{
    \begin{tabular}{p{14cm}}
        \centering 
        \resizebox{\linewidth}{!}{\includegraphics{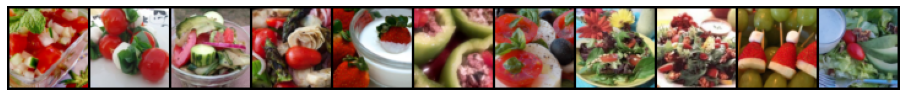}} \\
        \footnotesize \textbf{Image Query (Far Left Image)}: Shiraz Salad $\Longrightarrow$ \textbf{Recommendations}: Caprese On A Stick, Cooler Than A Cucumber Salad, Grilled Eggplant And Asparagus Salad, Casey And Leigh's Delightful Fruit Dip, ``Stuffed Bell Peppers, Greek Style'', Simple Caprese Salad, Fresh As A Daisy Spring Salad, Spinach And Hazelnut Salad With Strawberry Balsamic Vinaigrette, Grinch Kabobs, Gorgonzola Cheese Salad \\
        \resizebox{\linewidth}{!}{\includegraphics{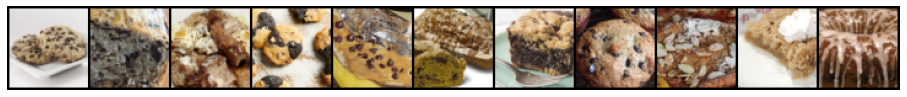}} \\
        \footnotesize \textbf{Image Query (Far Left Image)}: Chocolate Chip Cookies With Truvia® Baking Blend $\Longrightarrow$ \textbf{Recommendations}: {Bitter Chocolate, Lavender, And Banana Tea Loaf}, {Sugar-free French Toast Casserole}, {Chocolate Chip Crisscross Cookies}, {Peanut Butter Banana Melties}, {Pumpkin Chip Bread}, {Chocolate Chip Cookie Brownies}, {Easy Whole Wheat Banana Muffins}, {Eggless French Toast}, {Peanut Butter Apple Crisp}, {Sock It To Me Cake IV} \\
        \resizebox{\linewidth}{!}{\includegraphics{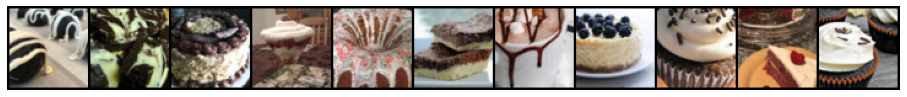}} \\
        \footnotesize \textbf{Image Query (Far Left Image)}: Cookie Balls $\Longrightarrow$ \textbf{Recommendations}: Cream Cheese Brownies III, Diane's German Chocolate Cake, Quick Lemon Chambord® Tiramisu, Mrs. Walker's Fruit Cake, Gooey Brownies With Shortbread Crust, Chocolate Bar Hot Chocolate, Instant Pot® Cheesecake, Buttercream Icing, Raspberry Cake Topped With Fruit, Chocolate Cupcakes With Caramel Frosting \\
        \resizebox{\linewidth}{!}{\includegraphics{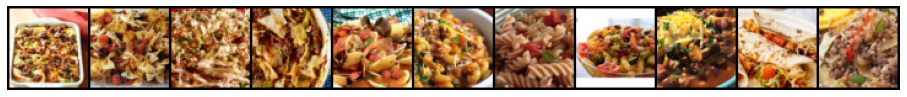}} \\
        \footnotesize \textbf{Image Query (Far Left Image)}: Sausage and Apple Breakfast Casserole $\Longrightarrow$ \textbf{Recommendations}: Fire And Ice Pasta, Cabbage Koora, Unstuffed Cabbage Roll, Clams And Chourico, One-pot Chili Mac And Cheese, Five Ingredient Pasta Toss, Simple Tasty Pasta Salad, Slow Cooker Taco Bean Soup, Southwest Bbq Chicken Tacos, Kimchi Bokeumbab (Kimchi Fried Rice) \\
        \resizebox{\linewidth}{!}{\includegraphics{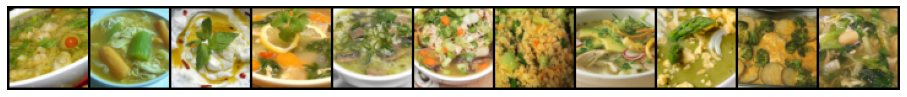}} \\
        \footnotesize \textbf{Image Query (Far Left Image)}: Meatball And Olive Stew (Albondigas Verdes) $\Longrightarrow$ \textbf{Recommendations}: Spicy Chicken Soup, Labneh (Lebanese Yogurt), Lemon Turkey Soup, Mexican Rice Soup With Mushrooms, Cambodian Lemongrass Chicken Soup, Savory Couscous Tabbouleh, Slow Cooker Chicken Pozole Blanco, ``Asparagus, Lemon, \& Mint Soup'', Easy Cheesy Chicken Bake, Chef John's Beans And Greens \\
        \resizebox{\linewidth}{!}{\includegraphics{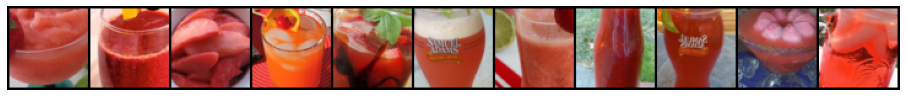}} \\ 
        \footnotesize \textbf{Image Query (Far Left Image)}: Frozen Strawberry Margarita $\Longrightarrow$ \textbf{Recommendations}: Mango Cherry Smoothie, Spiced Apples, Tropical Punch Cocktail, Strawberry Basil Balsamic Martini, Channel Marker, Strawberry Limeade,  Mississippi Sweet And Sour Barbeque Sauce, Barbie Shot, Incredible Punch, Cranberry Juice Surprise \\
    \end{tabular}
    }
    \caption{KG-VAE for Image-based Recommendation (Allrecipes.com). The KGE guidance helps to recommend more similar recipes to a given input image when compared to a (vanilla) VAE-based solution as shown in Figure \ref{fig:img-vae}.}
    \label{fig:img-based-rs}
\end{figure*}

\noindent \textit{Experimental Results.} Figure \ref{fig:img-based-rs} presents a few results for IRS. In each row, the far left image is the input image and the others are the recipes recommended. 
In overall, it is seen that KG-VAE finds similar recipes to a given recipe image. It, however, returns a few wrong recipes due to the texture of their images or the way they are cropped; for example, ``Labneh (Lebanese Yogurt)'' for ``Meatball and Olive Stew'' or ``Spiced Apples'' for ``Frozen Strawberry Margarita''. This may be alleviated by better cropping.

\section{Conclusion and Future Work}
\label{sec:conclusion-and-future-work}

We introduced the RECipe framework as a multi-purpose recommendation solution in which the core component is a multi-modal knowledge graph. We compared KGE models with the neural RS solutions for the (traditional) user-item recommendation task. We also presented that the pre-trained NLP embeddings enhance various applications such as zero-shot inference for
new users (or cold start problem) and conditional recommendation with respect to recipe clusters. We eventually demonstrated the application of RECipe for review-based and image-based recommendations. 

For future work, we would like to expand our knowledge graphs by adding other entity types such as nutritional facts for other purposes, e.g., nutritional fact prediction and ingredient replacement. For conditional recommendation (CR), we would like to investigate Gaussian Mixture Models (GMMs) to potentially assign every recipe to multiple clusters.

\begin{acks}

We appreciate the support that we received from Borislav Mavrin, Manasa Bharadwaj, and Yolanda Liu for accomplishing this research work.

\end{acks}

\bibliographystyle{ACM-Reference-Format}
\bibliography{acmart}


\newpage

\appendix

\renewcommand\thetable{\thesection.\arabic{table}}
\setcounter{table}{0}

\renewcommand\thefigure{\thesection.\arabic{figure}}
\setcounter{figure}{0}

\section{Knowledge Graphs}

\noindent 
The original datasets are very large but highly sparse which makes evaluating RS algorithm very hard. By following \citet{he2017neural}, we filter the datasets (or the KGs) by only keeping recipes and then users with at least $50$ and $10$ reviews, respectively. 
That ensures that we have more items than users in the KGs by filtering out non-popular items and isolated users.

\vspace{1mm}

\begin{table}[ht]
    \begin{subtable}{0.48\linewidth}
        \centering
        \begin{tabular}{|c|c|c|c|}
            \hline
            \multirow{2}{*}{Entity Type} & \multirow{2}{*}{Properties} & \multicolumn{2}{c|}{Num. Entities} \\ \cline{3-4}
            & & Original & Sample \\
            \hline
            Recipe & \begin{tabular}[c]{@{}c@{}} Name \\ Instructions \\ Keywords \end{tabular} & 265,812 & 2,417 \\ \hline
            Image & URL & 265,812 & 2,417 \\ \hline
            Category & Name & 289 & 125 \\ \hline
            Ingredient & Name & 3,182 & 685 \\ \hline
            Person & ID & 245,814 & 5,634 \\ \hline
            Review & Text & 1,325,716 & 149,161 \\
            \hline
        \end{tabular}
        \subcaption{Entity Types (Food.com)}
        \label{tab:foodcom-entities}
    \end{subtable}
    \begin{subtable}{0.48\linewidth}
        \centering
        \begin{tabular}{|c|c|c|}
            \hline
            \multirow{2}{*}{Relation Type} & \multicolumn{2}{c|}{Num.\ Triplets} \\ \cline{2-3}
            & Original & Sample \\
            \hline
            recipe:belongs to:category & 265,812 & 2,417\\
            recipe:contains:ingredient & 1,978,177 & 18,131 \\ \hline
            image:is-for:recipe & 265,812 & 2,417 \\
            image:belongs-to:category & 265,812 & 2,417 \\ 
            image:contains:ingredient & 1,978,177 & 18,131 \\ \hline
            ingredient:seen-with:ingredient & 463,138 & 29,442 \\ \hline
            person:posted:recipe & 265,812 & 2,417 \\
            person:likes:recipe & 1,241,281 & 149,161 \\ 
            person:wrote:review$^\dagger$ & 1,325,716 & 149,161 \\ \hline
            review:supports:recipe & 1,241,281 & 149,161 \\ \hline
        \end{tabular}
        \subcaption{Relation Types (Food.com)}
        \label{tab:foodcom-relations}
    \end{subtable}
    \caption{Food.com Entities and Relations \\ $^\dagger$ The original data include like, dislike, and neutral interactions. For that, the number of \textit{person:wrote:review} is greater than that of \textit{review:supports:recipe} in the original data.}
    \label{tab:foodcom}
\vspace{-4mm}    
\end{table}


\begin{table}[h]
    \begin{subtable}{0.48\linewidth}
        \centering
        \begin{tabular}{|c|c|c|c|}
            \hline
            \multirow{2}{*}{Entity Type} & \multirow{2}{*}{Properties} & \multicolumn{2}{c|}{Num. Entities} \\ \cline{3-4}
            & &  Original & Sample \\
            \hline
            Recipe  & \begin{tabular}[c]{@{}c@{}} Name \\ Instructions \end{tabular} & 49,698 & 4,883 \\ \hline
            Image & URL & 49,698 & 4,883 \\ \hline
            Ingredient & Name & 10,344 & 1,960 \\ \hline
            Person & ID & 1,160,267 & 8,702 \\ \hline
            Review & Text & 3,794,003 & 175,286 \\
            \hline
        \end{tabular}            
        \caption{Entity Types (Allrecipes.com)}
        \label{tab:allrecipes-entities}
    \end{subtable}
    \begin{subtable}{0.48\linewidth}
        \centering
        \begin{tabular}{|c|c|c|}
            \hline
            \multirow{2}{*}{Relation Type} & \multicolumn{2}{c|}{Num.\ Triplets} \\ \cline{2-3}
            & Original & Sample \\
            \hline
            recipe:contains:ingredient & 430,854 & 43,687 \\ \hline
            image:is-for:recipe & 49,698 & 4,883 \\
            image:contains:ingredient & 430,854 & 43,687 \\ \hline
            ingredient:seen-with:ingredient & 555,654 & 87,830 \\ \hline
            person:likes:recipe & 3,390,626 & 175,286 \\ 
            person:wrote:review$^\dagger$ & 3,794,003 & 175,286 \\ \hline
            review:supports:recipe & 3,390,626 & 175,286 \\ 
            \hline
        \end{tabular}
        \caption{Relation Types (Allrecipes.com)}
        \label{tab:allrecipes-relations}
    \end{subtable}
    \caption{Allrecipes.com Entities and Relations \\ $^\dagger$ The original data include like, dislike, and neutral interactions. For that, the number of \textit{person:wrote:review} is greater than that of \textit{review:supports:recipe} in the original data.}
    \label{tab:allrecipes}
\vspace{-6mm}    
\end{table}

\section{Why KG-VAE but not (Vanilla) VAE?}

\begin{figure*}[ht]
    \resizebox{\textwidth}{!}{
    \begin{tabular}{p{14cm}}
        \centering
        \resizebox{\linewidth}{!}{\includegraphics{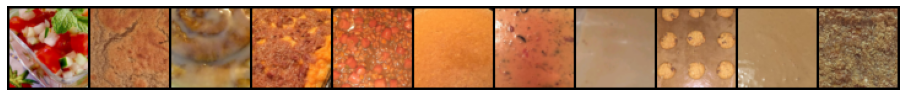}} \\
        \footnotesize \textbf{Image Query (Far Left Image)}: {Shiraz Salad} $\Longrightarrow$ \textbf{Recommendations}: {Caramel Sponge Pie}, {Cinnamon Roll Pancakes}, {Chef John's Sweet Potato Casserole}, {Beanie-weenie}, {Gluten Free Pumpkin Cheesecake}, {Junk Dip}, {Cocoa Coffee Milkshake}, {Cookies For Rookies}, {Curried Veggie Dip}, {Fresh Apricot Crisp} \\
        \resizebox{\linewidth}{!}{\includegraphics{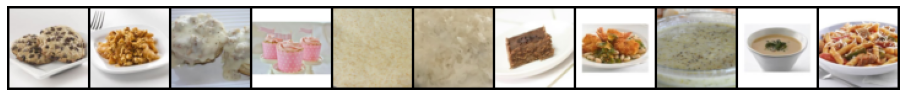}} \\
        \footnotesize \textbf{Image Query (Far Left Image)}: {Chocolate Chip Cookies With Truvia® Baking Blend} $\Longrightarrow$ \textbf{Recommendations}: {Apple Crisp With Truvia® Natural Sweetener}, {Sausage Gravy III}, {Aunt Kate's Strawberry Cake}, {Whole Wheat Pizza Dough Secret Family Recipe}, {Coconut Milk Cake Mix Cake}, {Lower Fat Fudge Brownies}, {Angry Shrimp With Tuscan White Beans}, {Low Fat Full Flavor Cream Of Broccoli Soup}, {Tuscan White Bean Soup}, {Zesty Penne, Sausage And Peppers} \\        
        \resizebox{\linewidth}{!}{\includegraphics{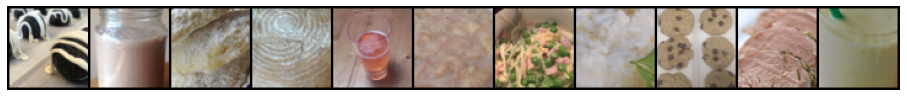}} \\
        \footnotesize \textbf{Image Query (Far Left Image)}: {Cookie Balls} $\Longrightarrow$ \textbf{Recommendations}: {Dairy-free Chocolate Peanut Banana Smoothie}, {Coconut Fingers}, {Multigrain Bread}, {Monaco}, {Banana Gravy}, {Tortellini Carbonara}, {Garlic Mashed Potatoes In The Slow Cooker}, {Oatmeal Chocolate Chip Cookies II}, {Simple Savory Pork Roast}, {Spiced Mango Lassi} \\
        \resizebox{\linewidth}{!}{\includegraphics{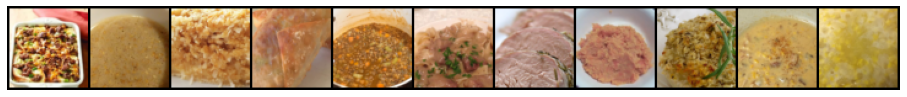}} \\
        \footnotesize \textbf{Image Query (Far Left Image)}: {Sausage And Apple Breakfast Casserole} $\Longrightarrow$ \textbf{Recommendations}: {Oriental Sesame Dip}, {Oatmeal Pie III}, {Cheeseburger Egg Rolls}, {Spicy Lentil Soup}, {Cindy's Beef Tips}, {Simple Savory Pork Roast}, {Carrot-sweet Potato Mash}, {Sweet And Sour Onions}, {Corn Chowder IV}, {Corn And Rice Medley} \\
        \resizebox{\linewidth}{!}{\includegraphics{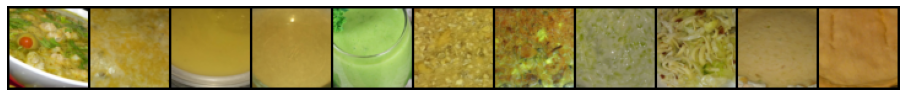}} \\
        \textbf{Image Query (Far Left Image)}: {Meatball And Olive Stew (albondigas Verdes)} $\Longrightarrow$ \textbf{Recommendations}: {Mexican Beef And Corn Casserole From Country Crock}, {Winter Squash Soup With A Sweet Heat}, {Dashi Stock (konbudashi)}, {Kale Banana Smoothie}, {Yam Casserole}, {Mom's Squash Casserole}, {Tzatziki I}, {Fried Cabbage And Noodles}, {Zucchini Cream Pie}, {Easter Pie} \\
        \resizebox{\linewidth}{!}{\includegraphics{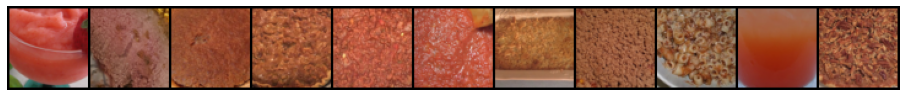}} \\    
        \textbf{Image Query (Far Left Image)}: {Frozen Strawberry Margarita} $\Longrightarrow$ \textbf{Recommendations}: {Herb Crusted Chuck Roast}, {Dixie Pie}, {Toll House Walnut Pie}, {Slow Cooker Ground Beef Barbecue}, {Canning Pizza Or Spaghetti Sauce From Fresh Tomatoes}, {Pineapple Casserole II}, {Strawberry-mango Pie}, {Lena's Pasta Fazul}, {Pineapple Upside-down Cake Martinis}, {German Chocolate Pecan Pie} \\
    \end{tabular}
    }
    \caption{Vanilla VAE for Image-based Recommendation (Allrecipes.com)}
    \label{fig:img-vae}
\end{figure*}

Our experiments show that the VAE results are not as good as those obtained by KG-VAE, as shown in Figure \ref{fig:img-vae}, for image-based recommendation. This can be explained by due to the fact that VAE is data-hungry and the (filtered) Allrecipes.com and Food.com KGs have only over 4.8K and 2.4K recipe images, respectively. On the other hand, KG-VAE improves the results, even with not many image examples, by receiving guidance signals from the KG entities. Besides, VAE tends to learn edges, colors, and textures to some extent whereas KG-VAE not only captures those features but also learns which recipes share similar ingredients.

\end{document}